\documentclass[10pt,journal,compsoc]{IEEEtran}
\IEEEoverridecommandlockouts

\ifCLASSOPTIONcompsoc
  \usepackage[nocompress]{cite}
\else
  \usepackage{cite}
\fi

\usepackage{lineno,hyperref}
\usepackage{amsmath}
\usepackage{amssymb}
\usepackage{listings}
\usepackage[]{algorithm2e}
\usepackage{textcomp}
\usepackage{graphicx}
\usepackage[pdf,tmpdir]{graphviz}
\usepackage[T1]{fontenc}
\usepackage{tikz}
\usepackage{pifont}
\usepackage{enumitem}
\begin{document}
\pagenumbering{gobble}% Remove page numbers (and reset to 1)

\onecolumn
\thispagestyle{empty}
\copyright~2020 IEEE. Personal use of this material is permitted. Permission from IEEE must be obtained for all other uses, in any current or future media, including reprinting/republishing this material for advertising or promotional purposes, creating new collective works, for resale or redistribution to servers or lists, or reuse of any copyrighted component of this work in other works.

\twocolumn
\pagenumbering{arabic}
\title{Semantic Access Control for Privacy Management of Personal Sensing in Smart Cities}

\author{Micha\l{}~Drozdowicz, 
		Maria~Ganzha, 
		Marcin~Paprzycki% <-this % stops a space
\IEEEcompsocitemizethanks{\IEEEcompsocthanksitem M. Drozdowicz, M. Ganzha and M. Paprzycki 
were with Systems Research Institute, Polish Academy of Sciences, Warsaw, Poland.\protect\\
Corresponding author: michal.drozdowicz@ibspan.waw.pl
\IEEEcompsocthanksitem M. Ganzha was with Department of Mathematics and Information Science, 
Warsaw University of Technology, Warsaw, Poland.% <-this % stops an unwanted space
%\IEEEcompsocthanksitem M. Paprzycki was with Warsaw Management Academy, Warsaw, Poland.
}}% <-this % stops an unwanted space

\IEEEtitleabstractindextext{%
\begin{abstract}
Personal and home sensors generate valuable information that could be used in Smart Cities. Unfortunately, typically, this data is locked out and used only by application/system developers. While vendors are partially to blame, one should consider also the ``binary nature'' of data access. Specifically, either owner has full control over her data (e.g. in a ``closed system''), or she completely looses control, when the data is ``opened''. In this context, we propose, a semantic technologies-based, authorization and privacy control framework that enables user to maintain flexible, yet manageable data access control policies. The proposed approach is described in detail, including implementation and testing.
\end{abstract}

\begin{IEEEkeywords}
Access control, semantic technologies, Smart City, data privacy, sensors, XACML, Attribute Based Access Control
\end{IEEEkeywords}
}

% \IEEEpubid{\copyright~2020 IEEE. Personal use of this material is permitted. Permission from IEEE must be obtained for all other uses, in any current or future media, including reprinting/republishing this material for advertising or promotional purposes, creating new collective works, for resale or redistribution to servers or lists, or reuse of any copyrighted component of this work in other works.}

\maketitle

\IEEEdisplaynontitleabstractindextext

\IEEEraisesectionheading{\section{Introduction}}


\IEEEPARstart{M}{any} Smart City projects rely on information collected from public sensor networks monitoring, among others: traffic, parking availability, pollution, noise, etc. (see, for instance,~\cite{iot_santander}). Examples, such as the city of Barcelona~\cite{iot_barcelona}, show benefits of such knowledge in governing the city, optimizing operational expenditures, and improving citizens' quality of life. However, the initial costs of such initiatives are very high, with major contributing factors including the purchase and installation of sensors, infrastructure cost (e.g. high-throughput network), or development and integration of software. Moreover, introducing new ``data sources'' often requires deploying new sensor networks, or upgrading existing ones (both generating substantial costs). Finally, to achieve the expected benefits, the ecosystem must be maintained and adapted to follow changes in technology and development and growth of the city.

Some of those shortcomings can be addressed by taking advantage of the rapidly growing number of personal, and home-based, IoT devices, therefore reducing the costs of hardware infrastructure needed to gather data. Moreover, the variety of citizen-owned sensing devices is systematically increasing, generating new dimensions of useful knowledge. For instance, the popularity of fitness tracking solutions has lead to a massive growth of health and lifestyle related data, which could be used to improve the medical and living conditions of the society.

Obviously, motivating the citizens to share their data ``with the city'' is a serious challenge, but it has been shown~\cite{klasnja2009exploring} that one of the key obstacles to achieving this is ``privacy management''. Specifically, how to facilitate adequate control over personal data, and thus convincingly assure the protection of privacy. Therefore, a successful solution, gathering personal sensor information for public use, should provide solid means of managing privacy preferences and fine-grained access control. Here, let us note that while there exist ways of anonymizing data to make it less sensitive, research on removing user/data anonymity limits the relevance of such approaches~\cite{el2011systematic,porter2008constitutional}. 

Furthermore, as discussed in~\cite{smart_city_privacy}, when dealing with health-related data and/or movement patterns extracted by fitness trackers, the actual challenge concerns \textit{relative perception of privacy}. Specifically, it involves not only \textit{which data is to be shared}, but also \textit{with whom} and \textit{for what purpose}. 

Finally, let us consider access to personal data by government agencies, e.g. related to criminal investigations, or national security. Analysis by Nojeim et al.~\cite{nojeim2014_governmentaccess} shows that, from legal and practical perspective, existing regulations and tools fail to reconcile public security with basic human rights and legal regulations (e.g. the GDPR). Therefore, when facing access requests, businesses storing the personal information rely on own judgment and/or interests, while agencies resort to broad, uncontrolled, and often unnecessarily detailed, surveillance. 

The described problem is an instance of a more general topic of \textit{access control}. It requires defining rules of who is allowed to access data, the same way as a company defines who can access a specific area in a building. Access control is well studied, and many approaches to solving it have been suggested and implemented. Here, Access Control Lists, Role Based Access Control or Attribute Based Access Control mechanisms have been created to tackle generalization of user roles and resource groups, static and dynamic Separation of Duties, spatiotemporal authorization, etc. Acknowledging this, note that several specific aspects of privacy management in Smart Cities need to be addressed:
\begin{itemize}
	\item Access requester is, likely, an organization. Moreover, the structure of the organization is, often, not known up-front. Hence, representation of (hierarchical) organizational structure is needed. Obviously, question of identity verification arises, but it is out of scope of this contribution.

	\item Data request is completed on behalf of an external organization, e.g. local government, which should be allowed to access only some data. Hence, token-based authorization (such as OAuth) is not feasible.
	
	\item Considered data is often a series/stream of observations that should be abstracted to types/categories, to avoid authorizing them individually. However, due to differences between devices/services, enforcement of a common observation vocabulary is not likely. Hence, use of access control mechanisms that depend on a fixed set of ``scopes'' (e.g. OAuth), may be challenging.
	
	\item Potential (large) scale of social participation, coupled with heterogeneity of data gathering applications, brings interoperability challenges that also materialize in access authorization. Differences in data representation necessitate either (i) conversion of data to some common format, which may not be feasible from closed data ecosystems (e.g. commercial fitness trackers), or (ii) introduction of a mapping layer. This would also result in authorization decisions involving centralized rules and policies.
	
	\item Time and location of issuing the request may not be essential, however certain spatiotemporal data related to the accessed information may be of use.
	
	\item Legal access to data (e.g. governed by GDPR) must be allowed, while rigorously controlled. Corresponding policies should consider purpose of use, retention routines, type of information, etc.

\end{itemize}

In this context, in~\cite{jms} we have proposed a semantically-enriched authorization system for fine-grained control of data access. Here, we expand on the idea, focusing on \textit{if} and \textit{in what way} ontological modeling, and semantic reasoning, can help manage privacy preferences in participatory sensing, within Smart Cities. The proposed solution recognizes that different information may be perceived as more or less private, depending not only on the nature of data, purpose of collection, and requesting entity, but also on purely subjective criteria. This, coupled with semantic representation and processing of pertinent meta-data, enables individuals to precisely manage their data access permissions. Finally, we recognize that certain legal regulations should be enforced and prioritized over the individuals' personal preferences. In this context, let us describe the use case scenarios, which guides the remaining parts of the paper.

\section{Overview of the use case scenario} \label{intro_use_case}

As discussed in~\cite{fitness_smartcity}, fitness data, collected by users for health tracking, could be useful for Smart Cities' agencies (e.g. public health organizations). It may not encounter known problems in adopting participatory sensing (also known as crowdsensing~\cite{wiki_crowdsensing}), such as the need for incentives~\cite{participatory_incentives} and/or change of behavior. Therefore, the proposed use cases involve tracking of an individual's movement habits, when the data is generated either by a GPS, or a pedometer (e.g. in a smartphone, smartwatch, or smart-shoes).

The general use case, is that of Sally, who uses a smartphone and a fitness application for tracking her running and cycling workouts. Thus far, she has collected data for personal benefits and shared it with friends (using some application). However, she is considering participation in a program analyzing sport activities in her home city. The primary use case (UC1) concerns the local Health Center, wishing to investigate the workout and training habits of the citizens. The second entity interested in her data (UC2) is the Police, investigating a crime in a certain area, searching for potential witnesses. For UC1 Sally would like to specify (independently) what information, and at what level of detail, she will share. UC2 illustrates how the proposed system handles legal obligations, while providing sufficient control over what data may be accessed under what conditions.

To deliver the needed functionality, we will build upon the semantically enriched Attribute Based Access Control system, introduced in~\cite{ict,jms,aciids}, and expand it with a more detailed model of privacy preferences, as well as means of handling legal access control policies.

Thus, in Section~\ref{state_of_art}, we give an overview of the state-of-the-art of solutions for enforcing privacy in Smart Cities as well as access control solutions making use of semantic technologies. Further, in Section~\ref{solution}, we briefly describe the SXACML access control system and discuss how to design an ontology that can be used to manage privacy and trust in Smart City. Finally, in Section~\ref{use_case} we revisit our guiding scenarios, to show in detail how proposed system enables citizens to manage access to their personal data, using the proposed system. 

\section{Related work} \label{state_of_art}

\subsection{Privacy and access control}

Let us first look into the methods of general access control, in which authorization policies and rules are used to validate if a \textit{Subject} is permitted to perform an \textit{Action} on a \textit{Resource} in a certain request \textit{Context}. We purposefully focus on the decision process and leave out consideration of ``orthogonal aspects'', such as identification, authentication, or action tracing. 

Attribute Based Access Control (ABAC) provides the most flexible, and context aware, approach to authorization. Here, \textit{Subject}, \textit{Action}, \textit{Resource}, and \textit{Context} are described with sets of attribute values. Authorization decision is based on evaluation of policies, that specify conditions on the attributes. The most common implementation of ABAC is the eXtensible Access Control Markup Language (XACML\footnote{\url{http://docs.oasis-open.org/xacml/3.0/xacml-3.0-core-spec-os-en.html}};~\cite{hu2014guide}).

\begin{figure}[htbp]
	\begin{center}
		\includegraphics[width=0.8\columnwidth]{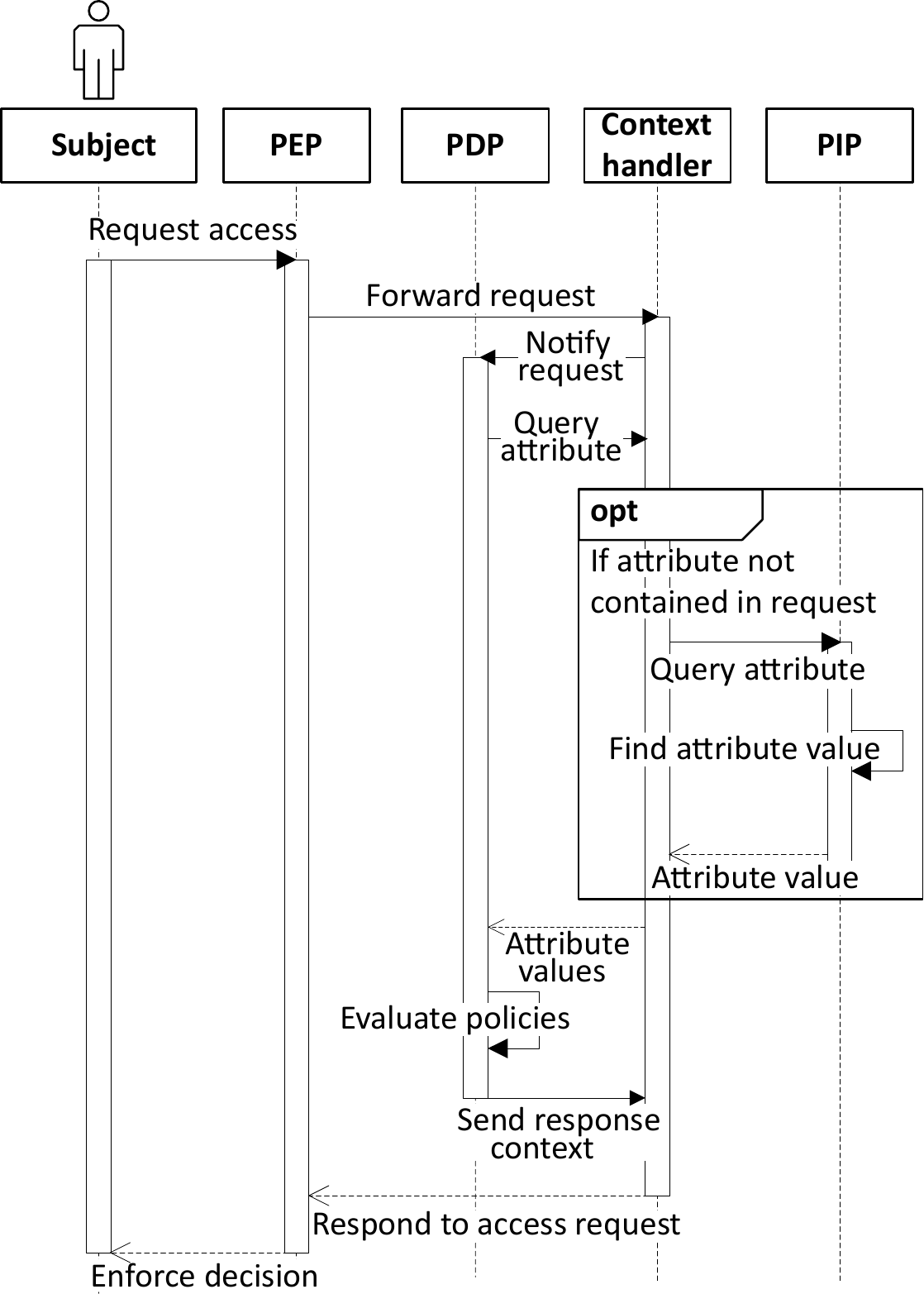}
		\caption{Evaluation of request in XACML}
		\label{fig:sxacml}
	\end{center}
\end{figure}

In Figure~\ref{fig:sxacml} we depict a typical sequence of actions undertaken during request evaluation, which follows the XACML standard.

When a system using XACML is configured, an administrator defines and manages policies within the Policy Administration Point (\textit{PAP}) and supplies them to the Policy Decision Point (\textit{PDP}). Once an access request is sent, by the \textit{Subject}, to the Policy Enforcement Point (\textit{PEP}), it is forwarded to the \textit{Context Handler} which, in turn, notifies the \textit{PDP}. Here, the \textit{PDP} verifies the values of all attributes used in the policy definitions. Such values may be found in the request itself, or need to be retrieved from one, or more, Policy Information Point (\textit{PIP}) components. Next, the \textit{PDP} evaluates policy rules, combines results (if multiple policies are involved) and builds the response context, which is returned to the \textit{Context Handler}. Results are then sent to the \textit{PEP}, which enforces the decision.

ABAC, however, omits (a)~implicit structure of entities requesting access, (b)~nature of data, or (c)~relations between \textit{Subject(s)} and \textit{Resource(s)}. This leads to significant complexity in defining and managing the policies. Regardless of improvements, like ALFA (Abbreviated Language For Authorization\footnote{\url{{http://docs.oasis-open.org/xacml/alfa-for-xacml/v1.0/alfa-for-xacml-v1.0.doc}}}),maintaining policies storing organizational and resource hierarchies, or more sophisticated relationships between data, requires major effort.

Additionally, authoring and administration of policies requires understanding of the policy definition language and the attribute space of the system under control. Therefore, ``policy management'' requires skilled specialists. As a result, ABAC has been adopted mostly in large organizations, e.g. in military, government, healthcare, or finance, where potential negative implications of unauthorized access justify expenditures related to managing policies. 

Note that our first use case assumes privacy control managed by the user who is not an access management expert. Therefore, it is crucial to provide tools and methods to easily express users' attitude towards data access.

Turning our attention back to the personal sensing,~\cite{klasnja2009exploring} explores the objections to participating in such programs, based on studying a group of people who used different sensors. It was observed that raised objections depend on what was recorded, in what circumstances, and on the created value. Moreover, giving users more knowledge and control over data increased the potential for adoption of crowdsensing-type approaches.

In this context, \cite{smart_city_privacy} proposes a framework for classifying citizen-related data, which also considers subjective \textit{feelings} about how personal the data is. For instance, location data could be seen as \textit{personal}, while ``generic traffic data'', not tied to an individual, would likely be \textit{impersonal}. Secondly, the \textit{purpose} of collecting data, ranging from ``service'' to ``surveillance'' is examined. For instance, use of location data for traffic management would, most probably, be seen as a ``service'', while using such information for predictive policing could be considered as ``surveillance''. When applying this framework to access control and privacy enforcement, two points are worth noticing. First, how ``personal'' a given ``piece of data'' appears to the user is rather complex. The data collected by any given sensor can be very personal if connected with personally identifiable information, or collected with high granularity (e.g. exact jogging location data). However, when such data is aggregated over time (e.g. distance run daily) and/or using strong anonymization (e.g. spatial k-anonymity;~\cite{spatial_k_anonymity}), it may be considered as not sensitive at all. Furthermore, different persons may care more, or less, about ``releasing'' personal data (see, information shared within social networks). Overall, ``level of sensitivity'' cannot be connected to a specific sensor. Instead, it is related to (a) type of observation, (b) its aggregation, (c) anonymization, and (d) ``personality'' of the user. Second, computer that evaluates access request has limited reasoning capacity. Therefore, it is important to let the user define permission rules, based (i) on the specific organization requesting the information and/or (ii) purpose of use, as specified in the request. For instance, users may assume that requests by the police department represent ``surveillance''. However, request for data to be used for traffic control, shifts the assessment towards ``service'' (as long as the user is willing to trust the police).

In~\cite{fitness_smartcity} and~\cite{clarke_jasist}, authors discuss the possibility of using personal health and fitness information. They also propose a privacy preserving architecture for data collection. Focus of their work is on aggregating and anonymizing the data, to become ``unbreakable'' by data mining. Here, let us note that numerous papers describe different solutions to data anonymization, especially for the location data; see, for instance~\cite{privacy_tesselation},~\cite{spatial_k_anonymity},~\cite{LIU2019421} and~\cite{LI20171}. However, these papers solve an issue that is, in a way, orthogonal to our concerns. We are interested in designing a system that gives its users the best possible control over their information, regardless of data anonymization. Hence, we accept that, in some cases, sharing fine-grained, personal information may be necessary. Moreover, we assume that the user might not want certain parties to acquire even anonymized information. Finally, we recall that some users may be willing to share ``very personal data'' regardless if it is anonymized or not.

Authors of~\cite{privacy_policies} propose a framework for managing and enforcing privacy policies in the context of data collection, as well as consent regulations, such as GDPR. The information model used in policies includes devices, entities (agents or organizations), data items, and purposes of use. Semantics of the policy elements can be expressed using subsumption of organizations and a partial order of attributes, which represents data item composition (e.g. street name as part of an address attribute). Each policy is a defined as a set of rules governing:
\begin{itemize}
	\item what attribute can be accessed, by which entity, under what condition (Data Communication Rule),
	\item for what purposes it can be used, and how long it can be retained (Data Usage Rule), and
	\item what are the rules for transferring the data to other entities, each governed by separate Data Communication and Usage rules.
\end{itemize}
Conditions within Data Communication Rules are defined using a formal, logics-based language, including negation and conjunction of predicates. Its relative simplicity brings the possibility of formal verification and provability of policies. 
The framework, however, offers limited possibilities to express semantics and relationships between the data attributes, which could lead to issues when working with more complex domains and policies. Additionally, given the scope of the framework, covering data collection, usage, and transfer, should the policies be fully enforced, the solution would need to be applied throughout the process of data collection, storage, and handling. This in turn may prove hard to implement in practice, given the wide variety of systems used by the data controllers and processors.

The Personalized Privacy Assistant Project\footnote{https://www.privacyassistant.org/}, by the Carnegie Melon University, seeks to provide individuals with tools enabling them to control how their data is collected and processed. Related publications such as~\cite{privacy_assistant_mobile1} and~\cite{privacy_assistant_mobile2} describe a mobile solution monitoring the device permissions (e.g. location, camera access) granted to various applications installed on the user's smartphone. It uses machine learning algorithms for clustering and classification, to group these programs into categories based on their functionality profile and purpose of data collection, finally assisting the user in making decisions about their privacy preferences. Additionally, to make configuration of preferences easier, the tool includes a number of privacy profiles and attempts to semi-automatically assign a user to one of them, based on answers to a simple survey. Compared to our research goals, the solution limits the scope of data under control to only the permissions recognized by the Android operating system and, therefore, does not address scenarios where the number of data items, or resources, is large or ever-changing. In a similar way, the set of applications installed on a mobile device do not change as often as the potential consumers of user data, in IoT and personal sensing scenarios.

Recent results of the project, described in~\cite{privacy_assistant_iot}, expand the solution to cover use cases dealing with a proliferation of IoT devices around the individuals. Here the authors address the problem that a typical person is continuously monitored by various devices, collecting personal information, and has little knowledge or control over the purpose and practices of data processing. They propose a distributed system, in which the personal assistant, residing on the user's smartphone, interacts with registries of surrounding IoT devices and uses its privacy preference policies to control what kind of consent it should give to device owners, i.e. data controllers. Finally, for certain types of more private data, external Policy Enforcement Points can be deployed that control what information can be provided to each data collector, depending on the privacy preferences of the data subject. The approach is sound, but tackles a somewhat different problem than the one we attempt to solve -- as in the case of~\cite{privacy_policies}, it deals primarily with user consent to collect and process data originating from devices owned and operated by third parties. In our context, we are more concerned with data generated by user-owned devices.

\subsection{Ontologies in Access Control} \label{onto_access_control}

In this context, let us note that knowledge representation and automatic reasoning, based on the structure and semantics of data, are dealt by ontology engineering. Over the years, it developed mature methods for formally representing concepts and their relationships. Here, an ontology is understood as a specification of a vocabulary for a domain, including classes of objects, relations, functions, and other concepts~\cite{GRUBER1993199}. Ontology-based models have been successfully applied in various areas, e.g. for genome modeling \cite{gene_ontology}; in healthcare (SAPPHIRE project;~\cite{semantic_web}), or in the Internet of Things \cite{interiot_onto}.
Overall, as shown in~\cite{ict,priebe,ferrini2009supporting}, by introducing semantic extensions to an ABAC system, it is possible to:
\begin{itemize}
 \item Define the structure of \textit{Subjects} and \textit{Resources} to closely model actual organizations and domains. Semantics also provides a consistent way of implementing RBAC and hierarchical resources.
 
 \item Represent additional relationships and reason over the model, to uncover implicit knowledge, to be used for checking request consistency, applicability of rules, and making decisions.
 
 \item Define mapping ontologies, making it straightforward to employ an ontological model of the domain that is reusable in other ways than just authorization.
 
 \item Flexibly and efficiently deal with heterogeneity, by utilizing ontology alignment and mapping.
 
 \item Define additional attributes that are automatically inferred from the information contained in ontologies.
 
 \item Delegate permissions from one \textit{Subject} to others in a hierarchy, by utilizing property transitivity.
 
 \item Infer conflicts between roles, by defining disjointness axioms in an ontology (thus, satisfying Separation of Duty). By defining rules, it is possible to tie the role assignment to dynamic conditions, such as time or the \textit{Action} being performed, to handle also Dynamic Separation of Duty.
 \end{itemize}

Another example of combining semantics and access control is~\cite{alcalde2007towards}, where authors propose SenTry -- a language and framework for personal privacy control. The solution is based on an OWL ontology modeling policies, and the Semantic Web Rule Language based (SWRL\footnote{\url{http://www.w3.org/Submission/2004/SUBM-SWRL-20040521/}}) rules for context-specific predicates used for decision making. Specifically, semantic reasoner evaluates applicable predicates, grouped into: filter, static and dynamic categories. This solution implements the ABAC model, but dismisses the de-facto standard of access control -- XACML. By building the solution completely from ground up, it misses the opportunity to benefit from the large number of existing (and, sometimes, very mature) tools built around XACML, dealing with handling of rule conflict resolution, request processing, geospatial functions, etc. 

Another solution, combining XACML with semantics is reported in~\cite{CHINGHSU201333}. Here, the LAPAR engine uses XML transformations (implemented as XSLT templates) to convert XACML policies to SWRL rules and further transform OWL ontologies and SWRL rules into Jess\footnote{\url{https://www.jessrules.com/}} inference engine statements. Proposed system reasons over combined knowledge, and computes authorization decisions. While presented results concern access to documents in a university, it is not clear (and somewhat doubtful) if the solution is capable of transforming the entire grammars of XACML, OWL, and SWRL into Jess rules solely by processing their XML representations. In absence of other use cases proving the concept also for other domains, we are not convinced the approach is applicable within the context of privacy control.

Summarizing, while the ABAC approach forms the best base for authorization systems in large-scale IoT applications, it lacks flexibility and meta-data modelling capabilities necessary in dynamic, heterogenous environments. Combining ABAC with semantic reasoning on ontological knowledge bases addresses these shortcomings. Finally, extending an established standard like XACML, instead of developing a purely semantic solution, brings numerous benefits, as well as giving the possibility to mix ontological and traditional ABAC-based rules in the same policy set. Let us now pursue this line of reasoning further.

\section{Ontologies for privacy management in Smart City} \label{solution}
\subsection{Semantic XACML}
In this context, in~\cite{ict,jms}, we have introduced a semantics-driven implementation of the \textit{PIP}, thus defining the Semantic XACML~(SXACML\footnote{\url{https://github.com/mdrozdo/SXACML}}) approach. The complete solution extends the XACML architecture in the following ways:
\begin{itemize}
	\item In addition to managing XACML policies, the \textit{PAP} module has been complemented with means of administering the ontologies used in the system. It includes a graphical front-end allowing one to define class mappings, expressions, and instances that are then added to the ontology and used during policy processing.
	
	\item The \textit{PDP} loads an additional resource finder module that handles multi-resource request scenarios, in which the access request does not specify a concrete resource, but rather a category that needs to be resolved to a set of individuals. The functionality of the semantic resource finder is depicted in Algorithm~\ref{alg:find_resources} and also handles resource class hierarchies, i.e. it can traverse an entire class-subclass structure defined in the ontology. Note that, in this scenario, the pure-XACML way of specifying the hierarchy relationships between resources, in policies, is very complex (see the XACML Hierarchical Resource Profile\footnote{\url{http://docs.oasis-open.org/xacml/3.0/rbac/v1.0/xacml-3.0-rbac-v1.0.html}}).
		
	\item A semantic \textit{PIP} module has been implemented, to enrich the set of attributes provided in the request with new information retrieved from the ontology. Thanks to the use of a semantic reasoner, the attribute values need not be explicitly defined in the knowledge base, but can also be inferred from other known facts. Additionally, we have introduced special attributes denoting the class identifier of each XACML attribute category (subject, resource, action, environment) that can be used in the policies for rules involving the type of a resource, or the role of the subject. As in the case of other attribute values, such classification of request categories can be deduced by means of automatic reasoning. The procedure of retrieving attribute values (labeled in Figure~\ref{fig:sxacml} as ``Find attribute value'') is performed according to Algorithm~\ref{alg:find_attribute})
\end{itemize}

\begin{algorithm}
	\SetKwInOut{Input}{input}\SetKwInOut{Output}{output}

	\Input{ URI of resource class $class$}
	\Output{set of permision decisions}	
	
	\BlankLine
	$O_{d}$ $\leftarrow$ load domain ontology\;
	$O_{m}$ $\leftarrow$ load mapping ontology()\;
	$O_{r}$ $\leftarrow$ new ontology \tcp*{temp request ontology}
	$O_{r}$ imports \{$O_{d}$, $O_{m}$\}\;
	run semantic reasoning on $O_{r}$\;
	\BlankLine
	$results$ $\leftarrow$ empty set of permission decisions\;
	$sol$ $\leftarrow$ query ontology for instances of $class$ \tcp*{takes into account entire subclass hierarchy}
	\ForEach{resource individual $I_{r}$ in $sol$}{
		$decision$ $\leftarrow$ evaluate policies for $I_{r}$
		add $decision$ to $results$
	}
	
	\KwRet{$result$}\;
	\BlankLine

	\caption{Evaluation for multiple resource class instances}\label{alg:find_resources}
\end{algorithm}

\begin{algorithm}
	\SetKwInOut{Input}{input}\SetKwInOut{Output}{output}

	\Input{evaluation context $ctx$, id of attribute to find $attrId$}
	\Output{bag of values of attribute}	
	
	\BlankLine
	$O_{d}$ $\leftarrow$ load domain ontology\;
	$O_{m}$ $\leftarrow$ load mapping ontology()\;
	$O_{r}$ $\leftarrow$ new ontology	\tcp*{temp request ontology}
	$O_{r}$ imports \{$O_{d}$, $O_{m}$\}\;
	
	\ForEach{category from $ctx$}{
		$I_{c}$ $\leftarrow$ new OWL individual\;
		\ForEach{attribute in category}{
			add property assertion to $I_{c}$\;
		}
		add $I_{c}$ to $O_{r}$\;
	}
	run semantic reasoning on $O_{r}$\;
	
	\BlankLine
	$result$ $\leftarrow$ empty bag of attribute values\;
	$sol$ $\leftarrow$ query ontology for attribute value\;
	\ForEach{result in $sol$}{
		$val$ $\leftarrow$ convert $result$ to an XACML attribute value\;
		add $val$ to $result$\;
	}
	\KwRet{$result$}\;
	
	\BlankLine
	\caption{Finding attribute values}\label{alg:find_attribute}
\end{algorithm}

The advantages of the SXACML approach include, but are not limited to:
\begin{enumerate}

\item Simplified policies -- information common to multiple policies can be ``extracted into the ontology'', resulting in the policies being represented in a ``more compact'' form.

\item Better support for RBAC -- role hierarchies can be modeled as ontology classes, user membership in a role can be inferred from attributes, and Separation of Duty can be verified by semantic reasoning.

\item More flexibility in defining relationships between concepts -- an attribute value may be inferred from a complex graph of linked data, utilizing properties of \textit{Subject}, \textit{Resource}, \textit{Action}, and \textit{Environment}.

\item Improved interoperability, by semantic mapping of disparate concepts in requests and policies -- allows decisions even in the case of different vocabularies.

\end{enumerate}

In prior work, we have used the semantic \textit{PIP} only as a provider of attribute values to policies (specified in XACML). The goal, in considered use cases, was to simplify XACML policies, and move domain models to OWL, assuming that the policy administrator has knowledge of XACML but not of OWL. We have also employed the OntoPlay\footnote{\url{https://github.com/mdrozdo/OntoPlay}} ontology editor~\cite{ontoplay, aciids} to assist the administrator in managing ontological concepts, such as \textit{Resource} categories, or \textit{Subject} roles. OntoPlay has proven to be a valuable tool enabling users not accustomed to semantic technologies to create complex class expressions and individual definitions, with applications not only in access control, but also in querying grid computational nodes at the University of Aizu, Japan~\cite{tsunami}. Furthermore, it has been utilized in student projects during the Semantic Technologies seminar at the Warsaw University of Technology, as well as in several master's theses defended at that institution, some of them resulting in publications, e.g. \cite{AIP_Chmiel, AIP_Szczekutek}.

However, in participatory sensing and personal privacy control, there is no system administrator -- the solution must be simple enough that the users are able to easily manage their own preferences and policies. We have, therefore, considered how to draw the boundary between OWL and XACML, taking into account that managing XACML policies manually is far beyond the capabilities of a casual user. Hence, we moved most responsibilities for the decision to the the semantic part, by defining the \texttt{PermittedRequest} class as a subclass of \texttt{Request} and providing the user with an OntoPlay-based interface that lets them define the relevant class expression in a point\&click manner, without knowledge of the ontology, and being fully agnostic of the XACML back-end. 

On the other hand, in the UC2 use case (i.e. the police investigation), access to the personal data should be rigorously controlled, considering legal conditions and obligations. Here, while it is possible to realize the ABAC model using semantics, it would introduce unnecessary complexity to the user. Moreover, implementing OWL concepts, capturing policy sets, combining multiple policies, obligations, etc., would change the policy processing engine. However, in comparison to subjective personal preferences, legal rules are likely to be relatively static, long-lived, and independent of the user. Therefore, legal policies can be implemented as standard XACML policies (by an access control expert).

In summary, we have separated (1) the subjective, dynamical personal privacy preferences -- defined in OWL -- and (2) the, potentially complex, static, legal rules defined as XACML policies and the policy sets. In the latter case, the seam between XACML and OWL follows earlier research -- the semantic \textit{PIP} infers and provides the \textit{PDP} values of certain attributes, and the remaining parts of policy processing are performed by the \textit{PDP} component. The final solution uses a policy combining algorithm to reconcile the privacy preferences with hard legal rules in the same policy set, giving higher priority to the regulatory requirements.

Note that, we only consider and describe the context of the access request permission and thus focus on the \textit{PAP}, \textit{PDP} and \textit{PIP}. Therefore, we purposefully ignore collecting and storing sensor/activity tracker data. We assume that, in a working system, another layer, responsible for data collection, would be instantiated. Examples of modules, realizing such functionality, can be found in~\cite{clarke_jasist, personal_data_vaults,fitness_integration}. Likewise, as mentioned earlier, we omit issues related to data anonymization. We assume that data requiring anonymization has already been processed, using techniques stated in Section~\ref{state_of_art}). Nevertheless, these simplifications do not influence the way that the proposed approach works. Finally, we leave out the details of the \textit{PEP} implementation, which would need to be tightly related to the way of storing the data as well as means of requesting the information by third parties. In the prototype implementation we have used WSO2 Identity Server\footnote{\url{https://wso2.com/identity-and-access-management}} as the gateway and \textit{PEP}.

Let us now turn our attention to another aspect of adapting SXACML to the Smart City use cases. Obviously, the crucial aspect of applying the system to a new domain is the selection or design of ontologies. While the term ontology has many definitions, here we understand it as formal representation of pertinent (application-specific) aspects of knowledge about a domain. Moreover, we have decided to use the Web Ontology Language (OWL\footnote{\url{https://www.w3.org/TR/owl2-overview/}}; \cite{owl_primer}) to formally represent ontologies.
One of the key features of OWL ontologies is that they can be reused by other ontologies, composed and adapted for more specific purposes. Hence, it is important to, first, search for existing resources in ontology catalogues such as the Linked Open Vocabularies\footnote{\url{http://lov.okfn.org/dataset/lov}}. However, we were unable to find a complete ontology covering the privacy management of data acquired from sensing devices. Therefore we have split the domain of interest into several parts, which are then combined into the final representation of the domain of interest.

To this effect, we discuss the following components of the ontological structure, used in the proposed solution:
\begin{itemize}
	\item Access Control ontology -- generic representation of ABAC concepts,
	\item Internet of Things ontology and Fitness Tracking ontology -- jointly representing \textit{Resources},
	\item Privacy ontology -- providing additional privacy-related concepts.
\end{itemize} 

Note that, following principles of ontology engineering, we have been re-using existing ontologies whenever possible, while modifying them only when necessary.

\subsection{Access Control ontology}

Let us start from the ontology describing concepts related to ABAC and XACML. This ontology has to be generic and domain independent. In~\cite{jms,ict} this purpose was fulfilled by a simplistic Request Ontology. Here, we introduce a more complete Access Control Ontology (ACO), as an ontological representation of the XACML request elements, but also providing core concepts related to data access. Basic elements of ACO are taken directly from the ABAC model, and reflect the same attribute categories as in XACML:
\begin{itemize}
  \item \texttt{Subject}
  \item \texttt{Resource}
  \item \texttt{Action}
  \item \texttt{Environment}
\end{itemize}
\noindent

For the \textit{Subject} part of the ABAC model, an ontology covering relationships between different organizational entities that can be authorized to access the personal data was needed. Hence, we have decided to directly use the W3C Organization Ontology\footnote{\url{http://www.w3.org/TR/2014/REC-vocab-org-20140116/}}. The XACML \textit{Subject} has been mapped to the \texttt{foaf:Agent} class, which may represent a person, group or organization. The ontology also contains concepts and relations needed to define complex organizational structures, and membership in them. Finally, we have reused the \texttt{org:Role} class, to be used in policies that assume decisions based on roles of the \textit{Subject} (in the RBAC approach). Moreover, we have defined classes related to the \textit{Resource}: 
\begin{itemize}
  \item \texttt{Sensitivity} -- capturing how personal the information is, and under what conditions it may be disclosed.
  \item \texttt{Confidentiality} -- describing the level of legal restrictions associated with the information.
  \item \texttt{Owner} -- specifying the entity (person or organization) owning the \textit{Resource} or being the main object described by the \textit{Resource}.
\end{itemize}

Another element is the \texttt{Trust} class, describing level of confidence of resource owner in given \textit{Subject}. While trust modeling is an interesting topic on its own, here, it is only a class, with subclasses corresponding to different degrees of confidence. Obviously, if needed, this class can be replaced by a more comprehensive ontology (fragment).

Finally, the ontology includes the \texttt{PurposeOfUse} class, describing the reason for requesting the \textit{Action} (how the obtained \textit{Resource} will be used).

\begin{figure}
	% \digraph[scale=0.3]{accesscontrol}{
	% 	node [	shape = box,
	% 		style = "rounded,filled",
	% 		color = steelblue,
    %         fontcolor = white,
	% 		fontname = "Arial" ];
	% 	edge [	fontname = "Arial" ];

	% 	"aco:Resource" -> "aco:Sensitivity" [label="aco:isSensitive"]
	% 	"aco:Request" -> "aco:Environment" [label="aco:inContextOf"]
	% 	"aco:Request" -> "aco:Resource" [label="aco:concernsResource"]
	% 	"aco:Subject" -> "aco:Trust" [label="aco:isTrusted"]
	% 	"aco:Request" -> "aco:Action" [label="aco:concernsAction";constraint=false]
	% 	"aco:Request" -> "aco:Subject" [label="aco:requestedBy"]
	% 	"aco:Resource" -> "aco:Confidentiality" [label="aco:isConfidential"]
	
	%     "aco:Trust" -> "aco:Action" [style="invis"]
	
	% 	"aco:Action" -> "aco:PurposeOfUse" [label="aco:requestedForPurpose"]
	% 	"aco:Action" ->  {"aco:Create","aco:Read","aco:Update","aco:Delete"} [label="has subclass"]
	%  }
	\includegraphics[width=1\columnwidth]{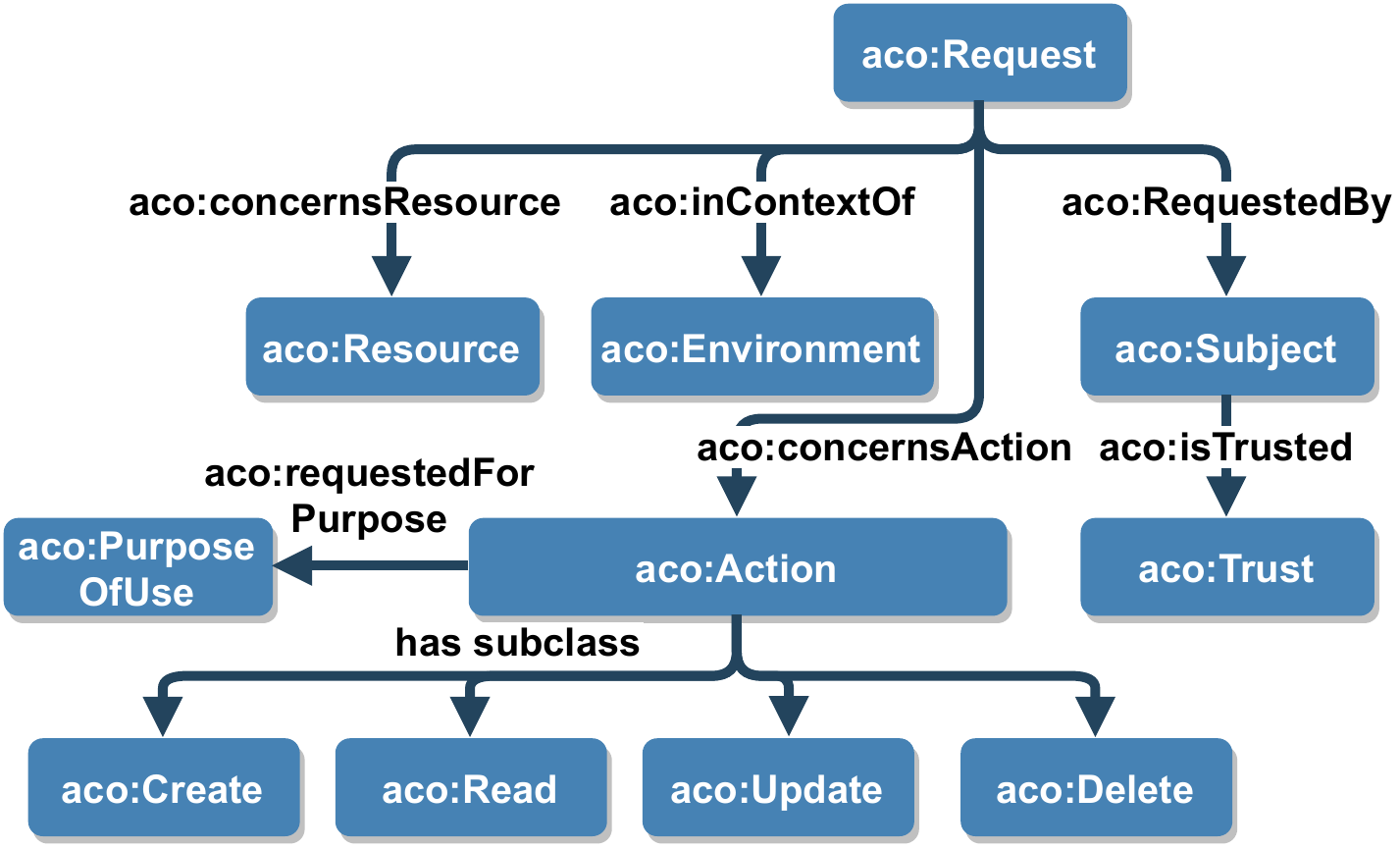}
	
	\caption{Access Control Ontology}\label{fig:access_control}
\end{figure}

Figure~\ref{fig:access_control} summarizes the Access Control Ontology.

\subsection{Domain ontologies} \label{sec:onto_iot}

First of all, considering that the guiding use cases deal with information collected using IoT devices, a ``sensor ontology'' is needed . This domain is well covered by the W3C Semantic Sensor Network Ontology (SSN\footnote{\url{https://www.w3.org/TR/2017/WD-vocab-ssn-20170105/}}), which contains vocabulary describing sensors, observations, and actuators, as well as observed properties, features of interest, etc. 

In SSN the \texttt{sosa:Observation} class represents a single act of measurement (\texttt{sosa:ObservableProperty}, e.g. heart rate) of a certain ``feature of interest'' (\path{sosa:FeatureOfInterest}, e.g. a specific person). It is described with (a) properties related to the sensor (\texttt{sosa:Sensor}, e.g. the heart rate monitor), (b) the feature/object that was measured, (c) the measurement procedure (\texttt{sosa:Procedure}, e.g. the method of measuring heart rate), and (d) detailed information about the result (\texttt{sosa:Result}). When accessing observations, no special requirements on what kind of actions can be performed are present (create, read, update or delete). We have extended the SSN ontology with an \texttt{AnonymizationProcedure} class (subclass of texttt{sosa:Procedure}) that describes the method used for removing personal information.

To represent privacy and access control in IoT scenarios, we needed to extend and adapt the SSN ontology to deal with accessing sensor generated observations, hence we have defined a mapping of \textit{Resource} and \textit{Action} from the ACO ontology to appropriate classes in SSN, as described in detail in Section~\ref{sec:mapping_onto}.

Second, to provide the needed vocabulary we have investigated several ontologies representing training activities and fitness tracking data. Here, the authors of~\cite{dragoni2018semantic} propose ontologies and a rule-based reasoner, for supporting people in following a healthy lifestyle. 
%As such the described ontologies cover both -- fitness activities and following dietary advice. 
The presented research focuses on eating habits and omits fitness activities, as well as data tracking, and thus is not a good fit for our needs. %Indeed, the ontology does not seem to contain any larger vocabulary of activities nor properties describing them.

In~\cite{dewabharata2013activity}, a framework for inferring person's physical state and activity, based on contextual information obtained from sensor network surrounding user, is proposed. Here, the \textit{Context Modelling Ontology} transforms external information into knowledge about user activity. Unfortunately, the vocabulary is rather high-level and is of limited use for our needs.

The knowledge base in the Physical Activity, Health and Fitness Knowledge Model~\cite{icfoods_physicalactivity} contains comprehensive vocabulary of physical activities. Furthermore, it includes concepts such as \textit{activity frequency}, \textit{activity intensity}, \textit{activity duration} and \textit{activity condition} (in terms of natural, social and legal environment). While some elements of this ontology, especially sports related, or describing workout intensity, could be reused, it is much too broad for our current needs. 

\begin{figure}[htpb]
	% \digraph[scale=.3]{fitnessTraining}{
	% 	node [	shape = box,
	% 		style = "rounded,filled",
	% 		color = steelblue,
    %         fontcolor = white,
	% 		fontname = "Arial" ];
	% 	edge [	fontname = "Arial" ];
		
	% 	"sosa:Observation" -> "fit:HealthMeasurement" [label="has subclass"]
	% 	"fit:IndoorRunning" -> "fit:TreadmillRunning" [label="has subclass"]
	% 	"fit:TrainingMetric" -> "fit:StepCount" [label="has subclass"]
	% 	"fit:Running" -> "fit:IndoorRunning" [label="has subclass"]
	% 	"fit:Cycling" -> "fit:OutdoorCycling" [label="has subclass"]
	% 	"fit:CardiovascularTraining" -> "fit:Cycling" [label="has subclass"]
	% 	"fit:HealthMeasurement" -> "fit:Weight" [label="has subclass"]
	% 	"fit:Training" -> "fit:HeartRate" [label="fit:measuredHeartRate"]
	% 	"fit:Training" -> "fit:TrainingMetric" [label="fit:describedWithMetric"]
	% 	"fit:HealthMeasurement" -> "fit:BloodPressure" [label="has subclass"]
	% 	"fit:HealthMeasurement" -> "fit:HeartRate" [label="has subclass"]
	% 	"fit:TrainingMetric" -> "fit:CaloriesBurned" [label="has subclass"]
	% 	"fit:TrainingMetric" -> "fit:Distance" [label="has subclass"]
	% 	"fit:Cycling" -> "fit:IndoorCycling" [label="has subclass"]
	% 	"fit:HealthMeasurement" -> "fit:Height" [label="has subclass"]
	% 	"fit:Training" -> "fit:CardiovascularTraining" [label="has subclass"]
	% 	"fit:HeartRate" -> "fit:Training" [label="fit:measuredDuring"]
	% 	"fit:CardiovascularTraining" -> "fit:Running" [label="has subclass"]

	%  }
	\includegraphics[width=1\columnwidth]{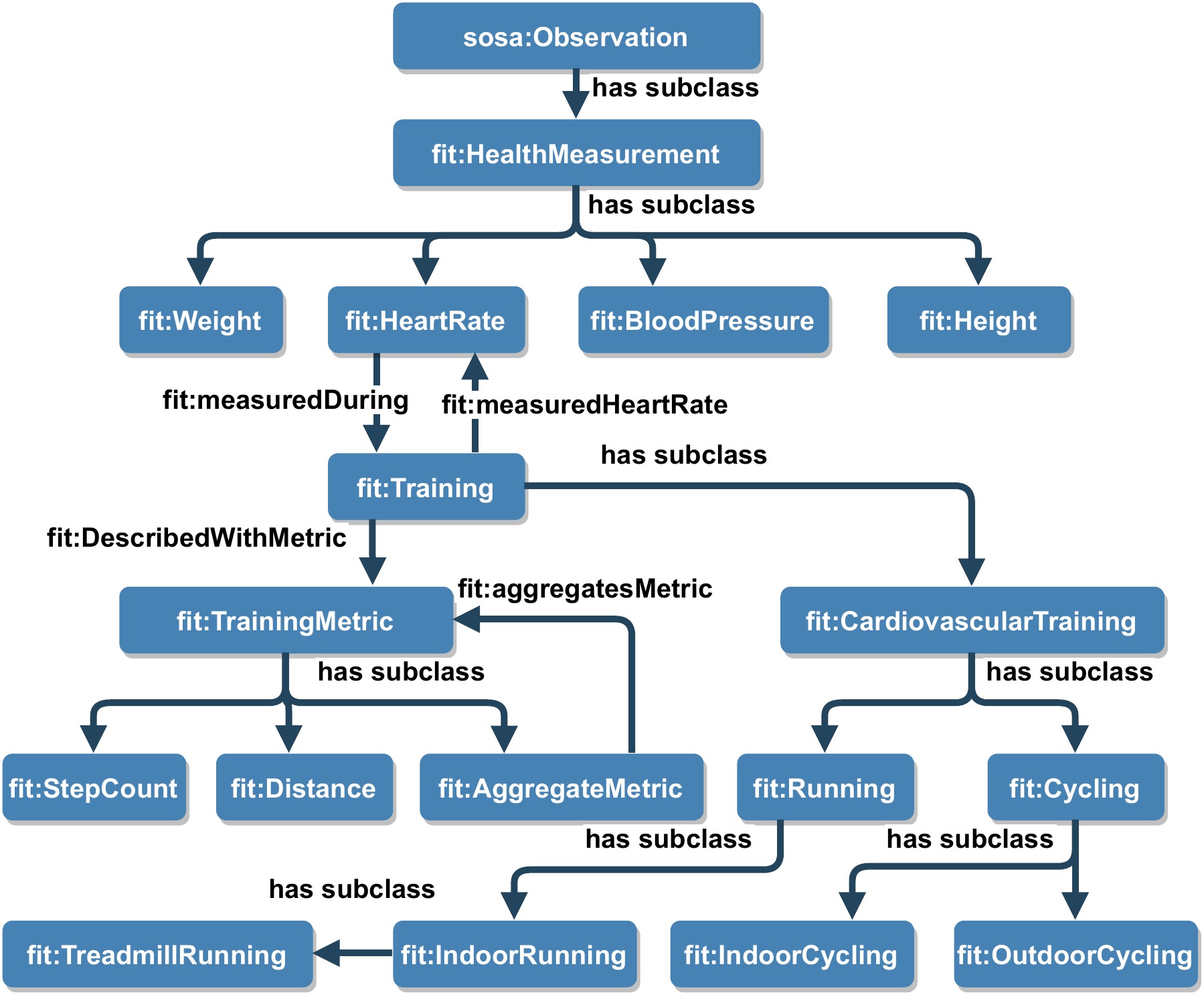}
	\caption{Fitness Tracking Ontology -- classes related to training types}\label{fig:fitness_training}
\end{figure}

\begin{figure}[htpb]
	% \digraph[scale=.35]{fitnessLocation}{
	% 	node [	shape = box,
	% 		style = "rounded,filled",
	% 		color = steelblue,
    %         fontcolor = white,
	% 		fontname = "Arial" ];
	% 	edge [	fontname = "Arial" ];
		
	% 	"sosa:Observation" -> "fit:Location" [label="has subclass"]
	% 	"fit:GeospatialMeasurement" -> "fit:Training" [label="fit:relatedToTraining"]
	% 	"fit:Training" -> "fit:GeospatialMeasurement" [label="fit:trainingLocation"]
	% 	"fit:GeospatialMeasurement" -> "fit:Route" [label="has subclass"]
	% 	"fit:Route" -> "fit:Distance" [label="fit:distance"]
	% 	"fit:Route" -> "fit:Location" [label="fit:includesLocation"]
	% 	"fit:Location" -> "fit:Route" [label="fit:partOfRoute"]
	% 	"fit:GeospatialMeasurement" -> "fit:Location" [label="has subclass"]
	% 	{rank="same"; "fit:GeospatialMeasurement"; "fit:Location"}
		
	%  }
	\includegraphics[width=1\columnwidth]{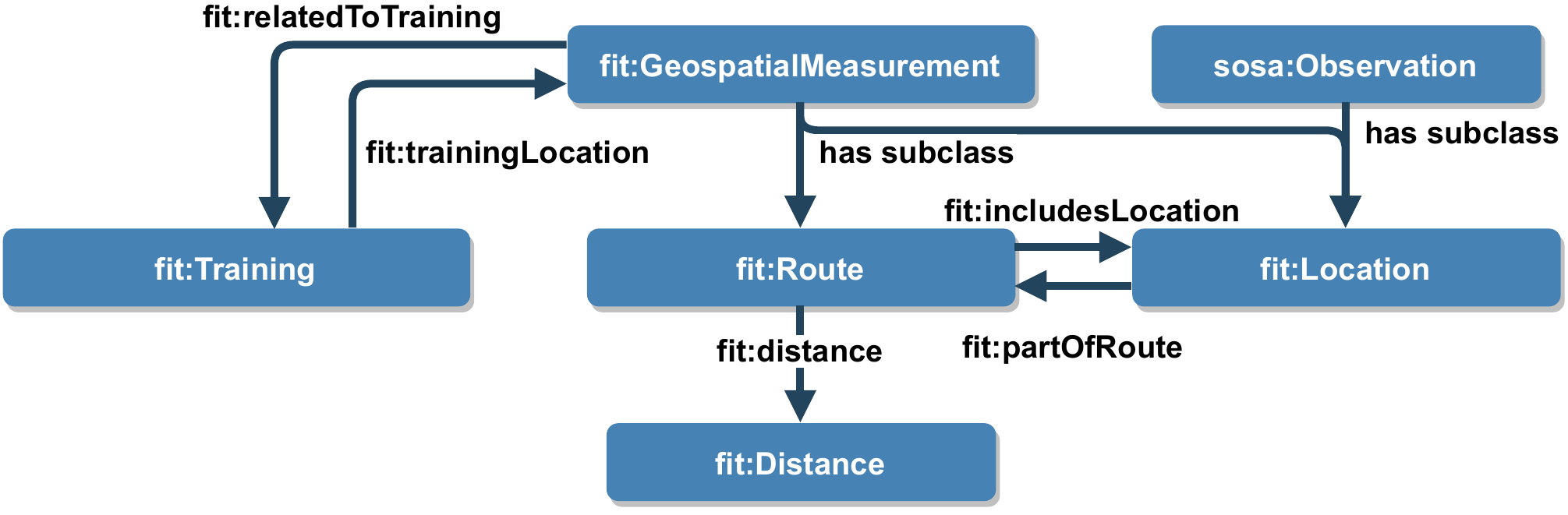}
	\caption{Fitness Tracking Ontology -- classes related to routes and locations}\label{fig:fitness_location}
\end{figure}

As a result of not finding an appropriate solution, we have decided to create a fitness tracking ontology, containing the most important elements relevant to collecting information to workouts and physical activities. To that effect, our Fitness Ontology imports the SSN ontology, and extends it with the following elements (also depicted in Diagrams~\ref{fig:fitness_training} and \ref{fig:fitness_location}). 
\begin{itemize}
	\item \texttt{Training} and its subclasses represent various workouts, e.g. running or cycling. This element is meant to be extended with a larger set of activities (perhaps using concepts from the Physical Activity Knowledge Base), depending on the application requirements.
	
	\item Several subclasses of the \texttt{sosa:Observation} class, representing the measurements related to training: (\texttt{BloodPressure}, \texttt{HeartRate}), or general physical metrics (\texttt{Height}, \texttt{Weight}).
	
	\item \texttt{TrainingMetric} representing workout attributes, such as calories burned, distance, or step count.
	
	\item \texttt{GeospatialMeasurement}, with subclasses \texttt{Location} and \texttt{Route}, capturing the training location.
\end{itemize}

The SSN and Fitness ontologies represent data that is to be subject to access control, i.e. the XACML \textit{Resource} category. The Access Control Ontology allows the \textit{Subject} category to be described as a person, machine agent, or an organization. Let us now consider how to describe the \textit{Action} category, taking into account attributes such as: purpose of use, retention policy, etc. 

When it comes to privacy ontologies, authors of~\cite{ppo} have introduced a lightweight ontology for privacy preferences, in the context of Semantic Web and linked data. Moreover, creators of Semantic Cyber Information Modeling Initiative (SCIMI\footnote{\url{http://privacyontology.org}}) have proposed a Domain Specific Language for describing a privacy meta-model. However, since 2015, there was no recognizable progress of this work.

The Platform for Privacy Preferences (P3P\footnote{\url{https://www.w3.org/TR/P3P/}}) is a specification, and an ontology, that allows web site authors to describe privacy practices in a machine readable format, enabling browsers to make semi-autonomous privacy related decisions. Even though the use case of P3P is different, developed ontology contains concepts useful in modeling privacy preferences in the participatory sensing, such as:
\begin{itemize}
  \item Classes capturing information describing web site visitors: name, email address, IP address, etc.
  \item Categories to classify information about a person: demographic, financial, health, location, etc.
  \item Purpose of use categories, such as: administration, contact, telemarketing, etc.
  \item Retention policies for the collected data.
\end{itemize}
The P3P specification has been retired in 2018. However, it is a good base for a privacy preference ontology.

In~\cite{hecker2009privacy}, an ontology, describing various aspects of privacy and their interrelationships is presented. The main goal was to categorize data privacy in certain situations (e.g. medical data of a patient admitted to a hospital). Rating is based on aggregating atomic scores, such as: \textit{Data Quality}, \textit{Security}, \textit{Data Subject's Rights}, \textit{Legitimate Grounds of Processing}, \textit{Transparency}, \textit{Consent}, \textit{Anonymity}. While this approach is quite interesting, it is not applicable to our use case as it does not take into account subjectivity of privacy. Specifically, even if policies and procedures are the same, some people may hesitate to expose personal data, while others do so without second thoughts.

The Privacy Preference Ontology (PPO), described in~\cite{sacco2011privacy}, is aimed at providing vocabulary and means of specifying privacy policies using RDF and SPARQL queries. The example uses FOAF to represent resources under control. Unfortunately, it captures only generic concepts (e.g. \texttt{PrivacyPreference}, \texttt{AccessSpace}, \texttt{Resource}, \texttt{Condition} etc.), which are already defined in our Access Control Ontology. 

Finally, authors of PrOnto~\cite{palmirani2018pronto} decided to model privacy and data protection concepts, in the context of GDPR, to enable legal reasoning and compliance verification. Therefore, it does not contain elements describing privacy preferences or data protection policies, but focuses on legal rules, rights, obligations, purpose of use, etc. Moreover, at the time of writing, no complete ontology could be found. Therefore it was hard to fully evaluate its suitability to our needs.

Eventually, we have decided that the P3P ontology can best serve as a base, however due to its size, for this report we have used only the elements relevant to our requirements, namely:
\begin{itemize}
	\item The hierarchy of subclasses of the \texttt{Data} class, representing various data categories. 
	\item The \texttt{Purpose} class as a representation of purpose of collection or use.
	\item The \texttt{Retention} class.
\end{itemize}

\subsection{Mapping ontology}\label{sec:mapping_onto}

Having described the ontological components representing the domain under consideration, let us consider in more details how they relate to each other. In order for the \textit{PIP} to access the knowledge base, it must first be consolidated in what we call the Mapping Ontology. It is built of several predefined mapping axioms, complemented with class expressions and / or individuals defined by the user as part of configuring their privacy preferences.
\begin{figure}[htpb]
	\includegraphics[width=1\columnwidth]{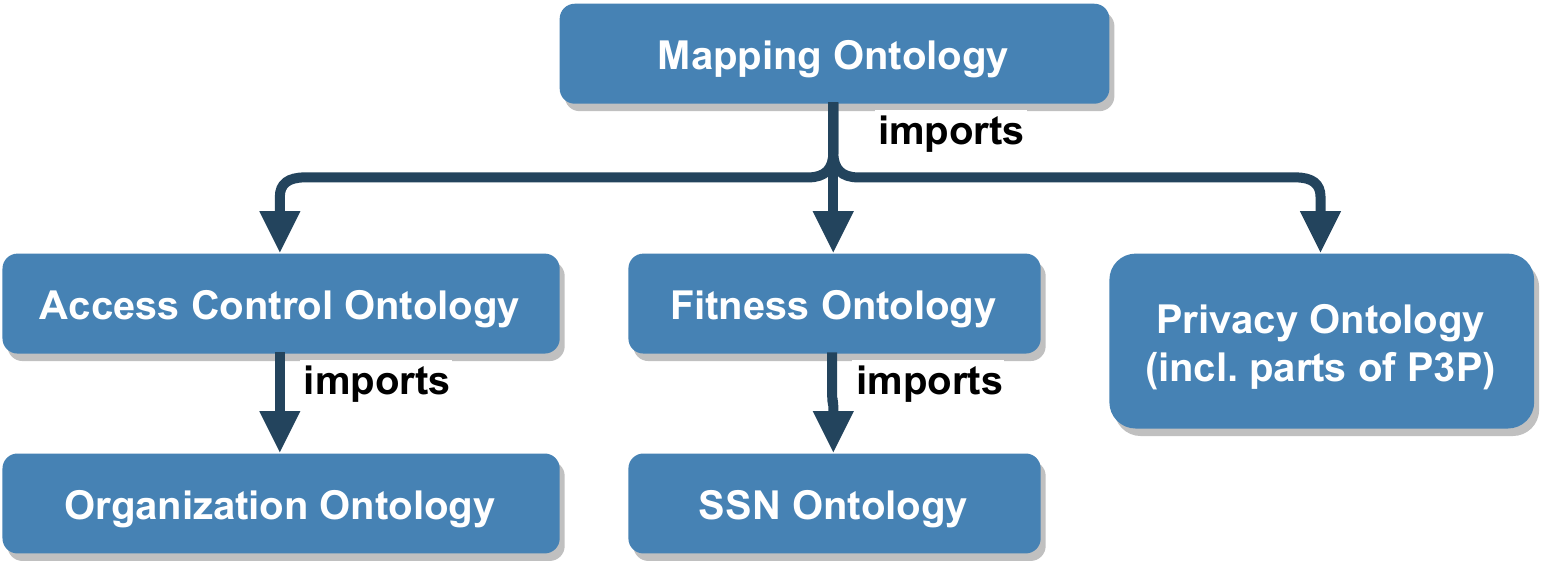}
	
	\caption{Imports hierarchy of ontologies}\label{diag:mapping}
\end{figure}

Figure~\ref{diag:mapping} depicts the high-level import hierarchy of the ontologies introduced in the previous sections. Specifically, the predefined mappings state that the \texttt{aco:Resource} class is a superclass of: \texttt{fit:Training}, \texttt{fit:TrainingMetric}, and \texttt{fit:GeospatialMeasurement}, which reflects what information could be requested. 

The mapping also joins the Fitness and Personal Privacy ontologies -- the \texttt{fit:HealthMeasurement} and \texttt{fit:GeospatialMeasurement} classes are marked as subclasses of \texttt{ppo:Health-data-category} and \texttt{ppo:Location-data-category} respectively. Moreover, we have added a custom category \texttt{ppo:FitnessData}, as a subclass of \texttt{ppo:OtherCategories}, that became a superclass for the \texttt{fit:Training} and \texttt{fit:TrainingMetric} classes. Finally, the \texttt{ppo:Purpose} class has been used as the range of the \texttt{ppo:hasPurposeOfUse} attribute, describing the \texttt{aco:Action} class (representation of the XACML \textit{Action} category). Analogously, we have added the \texttt{p3p:Retention} class as an attribute of \textit{aco:Action} (\texttt{ppo:hasRetentionPolicy}).

\section{Experimental Evaluation} \label{use_case}

With all elements in place, let us now illustrate how the proposed approach can be used in our two use case scenarios, introduced in Section~\ref{intro_use_case}: 
\begin{itemize}

	\item UC1: Health Center requesting aggregated (monthly) information about training metrics.

	\item UC2: Police department requesting Sally's locations during a specific time period.

\end{itemize}
\subsection{Health center}

\begin{figure}[htbp]
	\begin{center}
		\includegraphics[width=1\columnwidth]{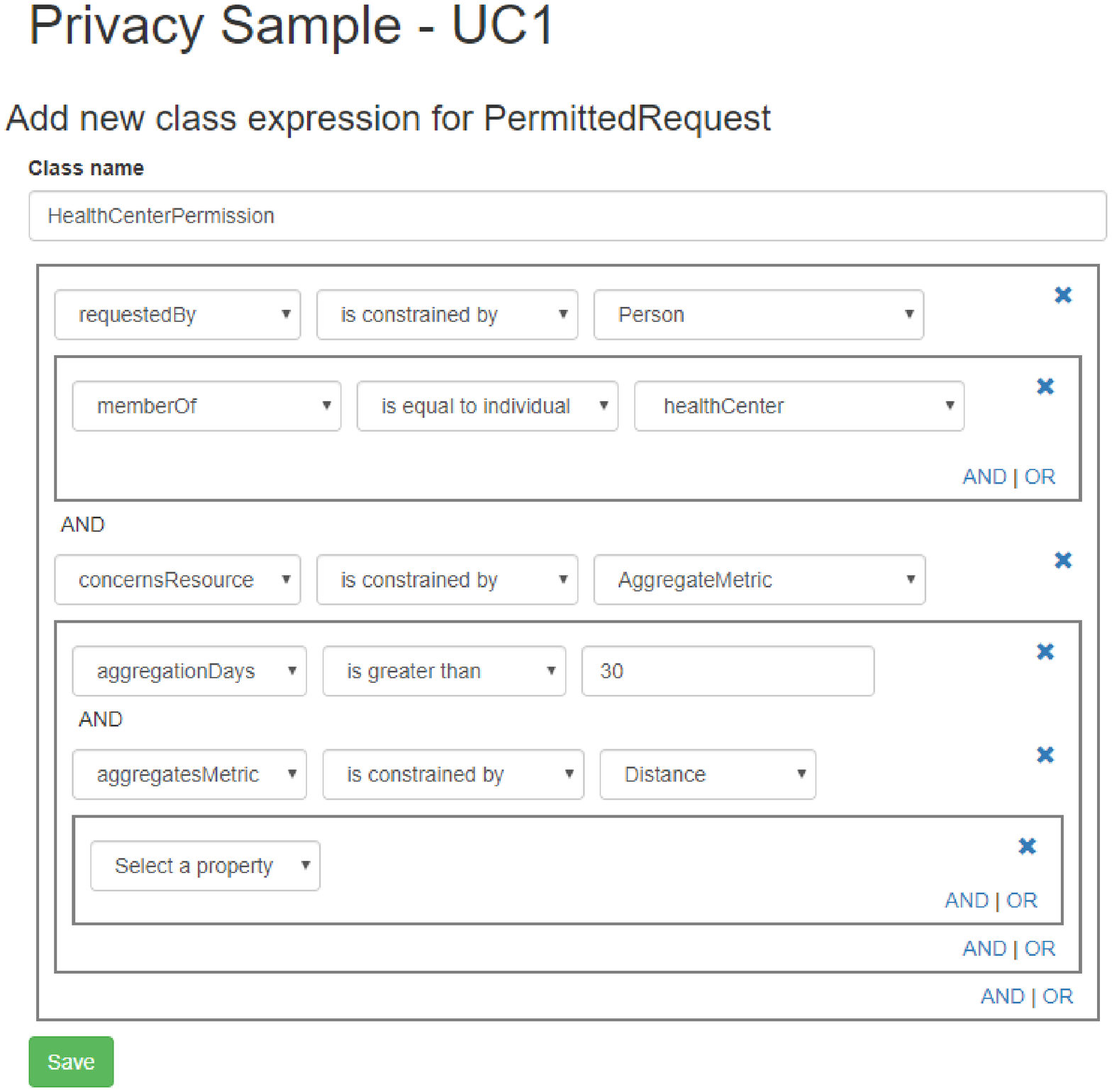}
		\caption{OntoPlay interface for Health Center permission}
		\label{fig:uc1_ontoplay}
	\end{center}
\end{figure}

We start with UC1, where the Health Center requests access to information about Sally's training metrics. First, Sally has to define her privacy preferences. Here, she specifies a permission stating that the requester, belonging to the Health Center organization, may access aggregated monthly distance observations. This policy can be easily created using OntoPlay, as depicted in Figure~\ref{fig:uc1_ontoplay}. The result is a class expression describing a subclass of \texttt{PermittedRequest} called \texttt{HealthCenterPermission} that is subsequently added to the ontology.

\begin{figure}[htbp]
	\begin{center}
		\includegraphics[width=1\columnwidth]{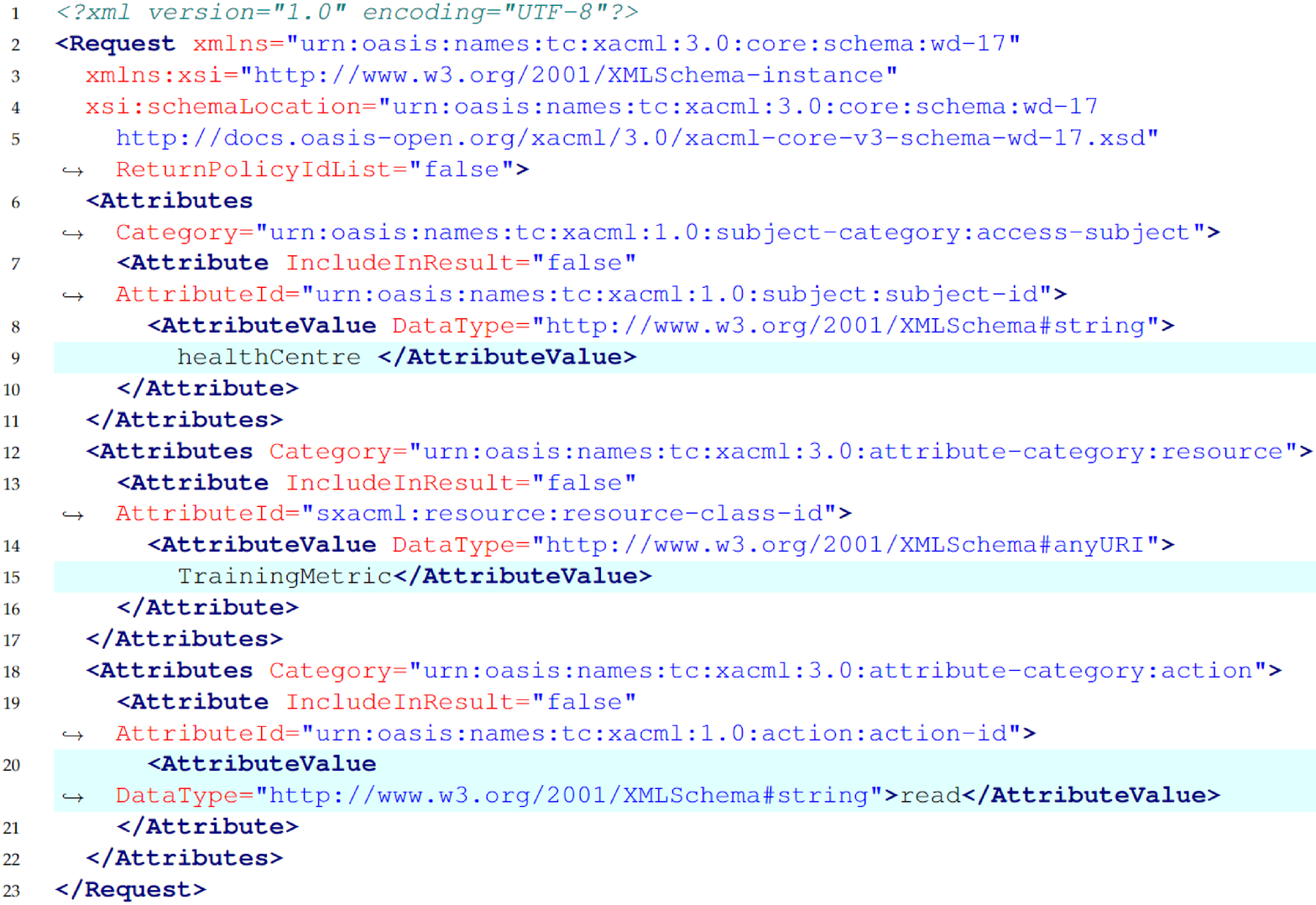}
		\caption{XACML request for UC1}
		\label{fig:uc1_req}
	\end{center}
\end{figure}

Figure~\ref{fig:uc1_req} presents the XACML request for the \texttt{Read} action (line 20), on resources of the \texttt{TrainingMetric} class (line 15), made by the Health Centre (line 9). Here, the semantic \textit{PIP} component adds a new individual of type \texttt{Request} to the temporary ontology.
As the request does not refer to a specific resource, but rather to a resource category -- the \textit{Resource} is only described as the \texttt{TrainingMetric} class. This is because the organization would like to collect as much relevant data as possible. Therefore, the initial step is to retrieve the appropriate resource individuals from the ontology. Considering the hierarchy of metrics shown in Figure~\ref{fig:fitness_training}, there exist several types of metrics: \texttt{StepCount}, \texttt{Distance}, and \texttt{AggregateMetrics}. Applying our semantic implementation of the XACML Hierarchical Resource Profile, the system translates this request into multiple decisions, based on the results of inference, which individuals in the ontology are instances of the \texttt{OutdoorTraining} class or its subclasses. The following evaluation steps are subsequently repeated for each resource.

The \textit{Subject} is specified with the class \texttt{HealthCentre}. Hence, an individual of that class is added to the request ontology, and linked to the request individual. The request individual is also connected to the \textit{Resource}. Next, the semantic reasoner can infer if the request individual satisfies all constraints specified by Sally, and classify it as \texttt{HealthCenterPermission}, and as \texttt{PermittedRequest}. Note that, the XACML policy includes a condition on the request class id attribute. Therefore the \textit{PIP} returns a collection of classes describing the request individual: \texttt{Request}, \texttt{PermittedRequest} and \texttt{HealthCenterPermission}. The rule evaluates to a \textit{true} value and the request is permitted by the \textit{PDP}.

\subsection{Police department}

\begin{figure}[htbp]
	\begin{center}
		\includegraphics[width=1\columnwidth]{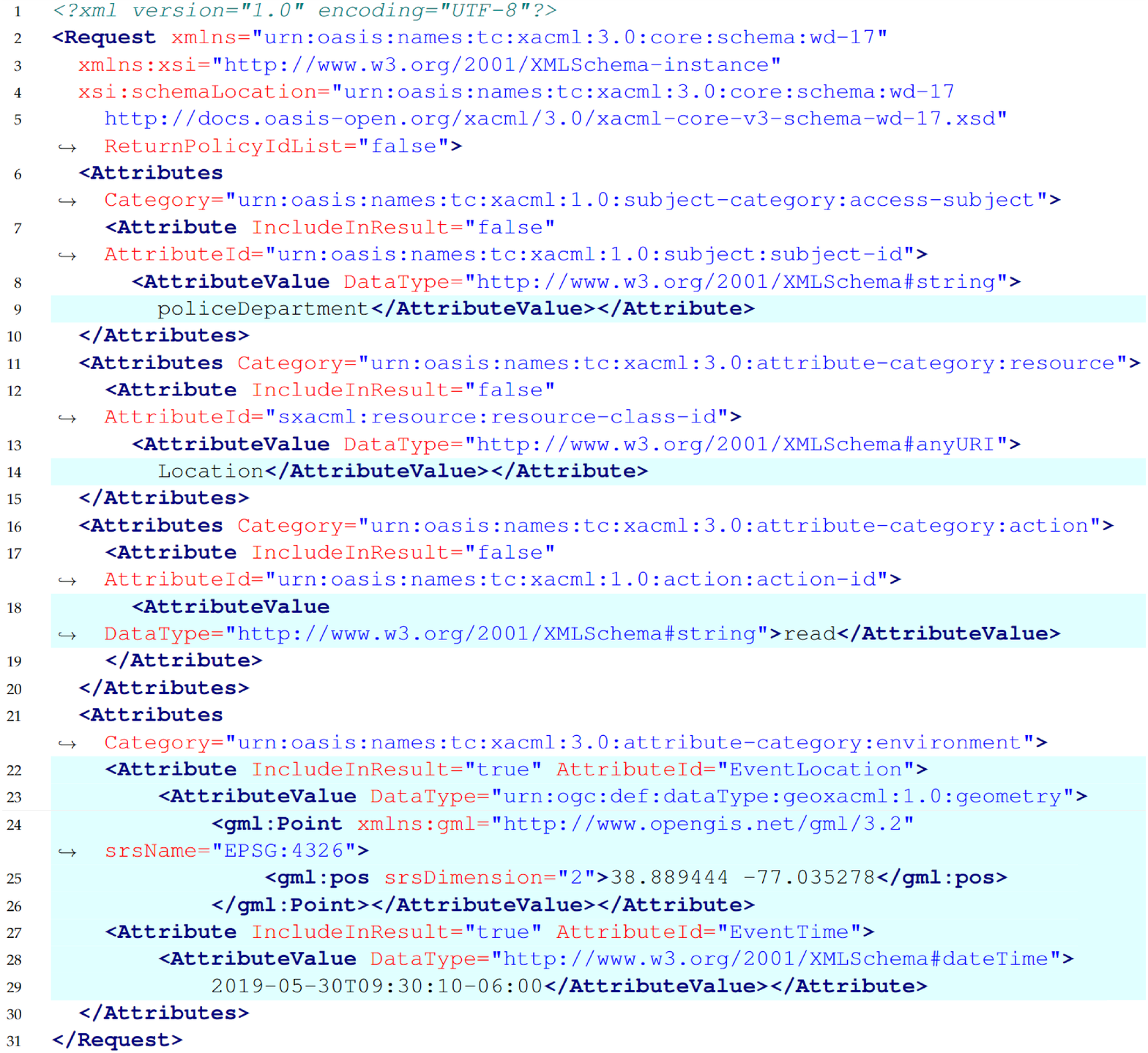}
		\caption{XACML request for UC2}
		\label{fig:uc2_req}
	\end{center}
\end{figure}

\begin{figure}[htbp]
	\begin{center}
		\includegraphics[width=1\columnwidth]{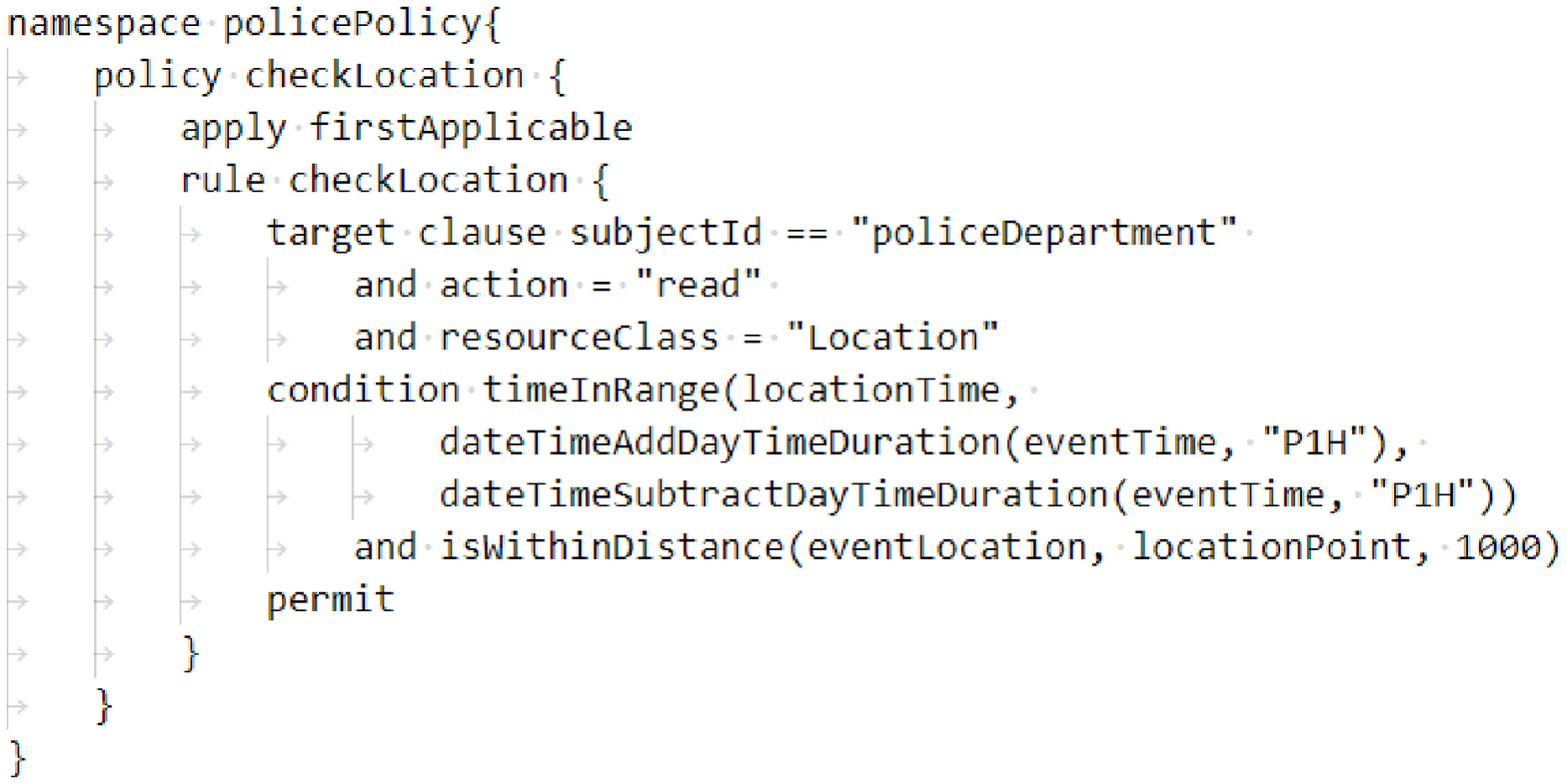}
		\caption{Policy for UC2}
		\label{fig:uc2_policy}
	\end{center}
\end{figure}

Next, let us consider the case of the Police Department, requesting location records from the night a crime took place (case UC2). Here, it is not up to the individual to define the rules of data access, as they represent en existing legal framework. In our, somewhat artificial, example we assume that the policy permits the Police to access information about the location of an individual as related to a committed crime. In the real-world, such a request would need to be accompanied by a warrant, i.e. specifying event location and time. Such warrant would need to be verified and digitally signed before being included in the data access request. Here, we will omit details as to how a warrant issuer can secure the request, and protect it from tampering. Nevertheless, let us stress that existing specifications such as the XACML XML Digital Signature Profile\footnote{\url{http://docs.oasis-open.org/xacml/3.0/dsig/v1.0/xacml-3.0-dsig-v1.0.html}}, provide appropriate solutions. The example request, depicted in Figure~\ref{fig:uc2_req}, contains the following attribute values: 
\begin{itemize}
	\item The \texttt{Subject} is Police Department (line 9).
	\item The \texttt{Resource} to be accessed is \texttt{Location} -- which should be understood not as a single, specific, position, but rather as all permitted locations (line 14).
	\item The \texttt{Action} is Read (line 18).
	\item \texttt{Environment} encloses attributes related to the crime event location and time (lines 22-29).
\end{itemize}
The associated policy is presented in Figure~\ref{fig:uc2_policy} (encoded in ALFA for brevity). It contains a number of conditions on different attributes. The \textit{Subject} is limited to the Police Department, and only \texttt{Read} actions are permitted. Moreover, the request is only permitted if the \textit{Resource} location attribute is within a one kilometer radius from the event location, and the position record time is in the period of one hour before and after the event. To apply geospatial comparison, the policy makes use of an extension to the XACML standard -- the Geospatial eXtensible Access Control Markup Language (GeoXACML\footnote{\url{https://www.opengeospatial.org/standards/geoxacml}}). Such spatiotemporal conditions would be hard to define in a typical OWL reasoner and would require specific geospatial extensions to the ontology language, as well as the inference engine. The policy could also be expanded to feature conditions related to the purpose of use of the information, warrant chain, etc. Additionally, not illustrated in the listing, the policy is also part of a Policy Set together with other policies (e.g. the one used in UC1), configured with a combining algorithm which secures that the law enforcement regulations override any user-defined preferences. 

In this case, the request defines the resource as an instance of the \texttt{Location} class and therefore the \textit{PDP}, again, needs to resolve it to a number of individual resources. The semantic reasoner is used by the resource finder to fetch concrete instances of the \texttt{Location} class from the ontology. While evaluating the policy for each of the locations, the \textit{PDP} does not encounter the \texttt{locationTime} and \texttt{locationPoint} attributes from the \texttt{Resource} category and therefore it requests their values from the (semantic) \textit{PIP}. Having acquired the attribute values, the policy condition is evaluated, by means of date-time XACML functions and the GeoXACML engine. The result is a multi-resource decision consisting of a number of individual responses, one for each location contained in the ontology.

\section{Concluding remarks}

In this paper we have considered how a semantically enriched Attribute Based Access Control system can be applied to (self-)management of user data privacy in Smart Cities. We have shown that practical application of semantic technologies brings important advantages for development of flexible, though robust, privacy controlling environments. In this context, our main contributions are as follows.

\begin{itemize}
	\item 	We have reviewed existing ontologies, covering different aspects of the domain of interest, including sensors, fitness tracking and personal privacy, finally composing well established vocabularies into a complete ontology. This outlines the path that should be followed when systems similar to ours are to be developed for other domains.
	
	\item On the basis of our earlier work, we have presented a more refined approach to combining the XACML policies with semantical reasoning. Here, among others, we have taken into account observation that certain rules may need to be more rigid than others. The proposed approach allows ``mixing and matching'' (depending on specific circumstances of the developed system) XACML rules with attributes resulting from semantic reasoning. In other words, the boundary between XACML and semantics can be instantiated as needed.
	
	\item We managed to separate the (fixed) ``rules of the system'' that are to be formulated by specialists, from user-preferences. Here, each user can (``dynamically'') formulate her/his rules, representing personal attitude towards privacy. User preferences will be captured within the system, without the need to change the rules.

	\item Expression of personal preferences does not require knowledge of semantic technologies. Rather, it is realized using OntoPlay, a novel interface to ontology-driven systems. 
	
	\item Moreover, use of OntoPlay allows easy modification of system ontology. After ontology is modified, it will automatically materialize in the user interface, without the need of changing the system code.
	
\end{itemize}

One area that we have not included in the research, but intend to investigate in the future, is the specification of obligations, i.e. additional actions that should be performed by the PEP, following the enforcement of the decision. The considered solution will fully support the default XACML obligation specification format, but taking advantage of the rich body of knowledge regarding semantic web services could improve the possibilities of describing mandatory data storage and processing regulations.

\bibliographystyle{IEEEtran}
\bibliography{sxacml}


\begin{IEEEbiography}[{\includegraphics[width=1in,height=1.25in,clip,keepaspectratio]{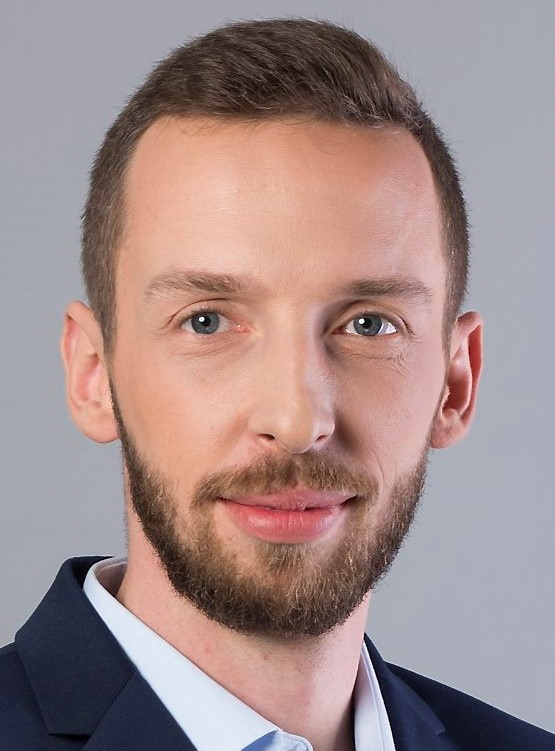}}]{Micha\l{} Drozdowicz}
obtained M.Eng. degree in Applied Computer Science from Warsaw University of Technology, Warsaw, Poland in 2007 and is currently pursuing Ph.D. at the Systems Research 
Institute, Polish Academy of Sciences. His research interests include semantic data processing, information privacy and distributed computing. 
\end{IEEEbiography}
	
\begin{IEEEbiography}[{\includegraphics[width=1in,height=1.25in,clip,keepaspectratio]{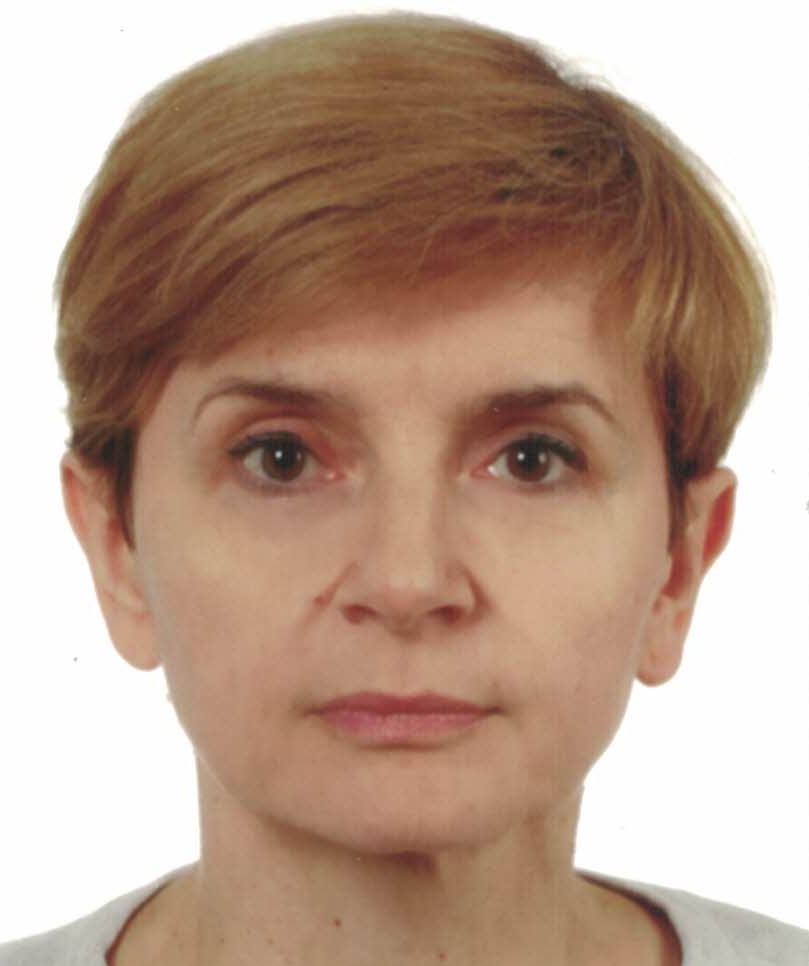}}]{Maria Ganzha}
is an Associate Professor in the Warsaw University of
Technology (Warsaw, Poland). She has an MS and a PhD degree in
mathematics from the Moscow State University, Russia, and a Doctor of
Science degree in Computer Science from the Polish Academy of Sciences.
Maria has published close to 200 research papers, is on editorial boards
of 5 journals and a book series, and was invited to program committees
of more than 250 conferences.
\end{IEEEbiography}
	
%\newpage
	
\begin{IEEEbiography}[{\includegraphics[width=1in,height=1.25in,clip,keepaspectratio]{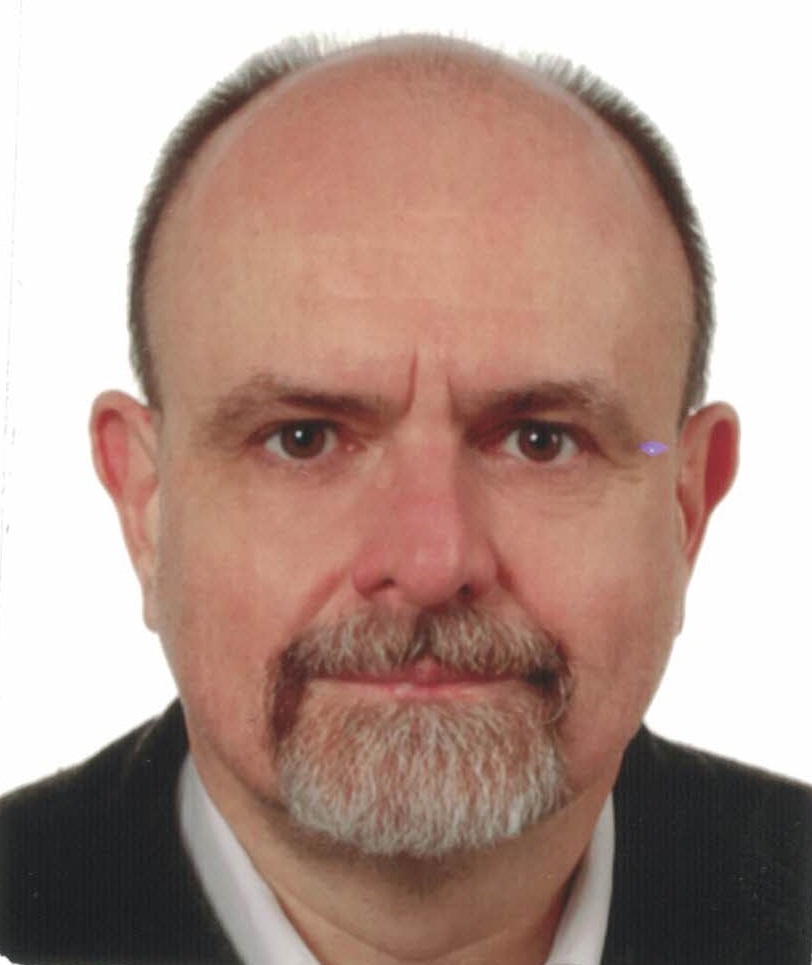}}]{Marcin Paprzycki}
	is an associate professor at the Systems Research
	Institute, Polish Academy of Sciences. He has an MS from Adam Mickiewicz
	University in Poznań, Poland, a PhD from Southern Methodist University
	in Dallas, Texas, and a Doctor of Science from the Bulgarian Academy of
	Sciences. He is a Senior Member of IEEE, a Senior Member of ACM, a
	Senior Fulbright Lecturer, and an IEEE Computer Society Distinguished
	Visitor. He has contributed to more than 500 publications and was
	invited to the program committees of over 800 international conferences.
\end{IEEEbiography}

\end{document}